\documentclass{PoS}
\usepackage{latexsym, amsmath, amssymb}

\providecommand{\abs}[1]{\lvert#1\rvert}    
\providecommand{\norm}[1]{\lVert#1\rVert}   
\providecommand{\txt}[1]{\mathrm{#1}}
\DeclareMathOperator{\hd}{h_0^{\dagger}}
\DeclareMathOperator{\h}{h_0^{\phantom{\dagger}}}
\DeclareMathOperator{\one}{1\hspace{-2.6pt}\mathsf{I}} 
\DeclareMathOperator{\e}{e}
\DeclareMathOperator{\arcosh}{arcosh}

\title{Spectral properties of the non-hermitian Wilson-Dirac operator in the
Schr{\"o}dinger functional}

\ShortTitle{Spectral properties of the non-hermitian Wilson-Dirac operator in the
SF}

\author{\vspace{2cm}Shinji Takeda, \speaker{Oliver Witzel} and Ulli Wolff\\%
        Humboldt Universit{\"a}t zu Berlin, Institut f{\"u}r Physik, Newtonstr. 15, 12489 Berlin, Germany\\
        E-mail: \email{witzel@physik.hu-berlin.de}\vspace{-4cm}\flushright
\begin{minipage}{2.5cm}
\textnormal{HU-EP-07/45 \newline SFB/CPP-07-60}
\end{minipage}\vspace{3cm}}

\abstract{We report on some preparatory investigations for the simulation of the QCD Schr{\"o}dinger functional with a non-hermitian 
polynomial hybrid Monte Carlo algorithm. The complex spectrum of the non-hermitean
free operator with SF boundary condititons is computed semianalytically. 
Then it is shown how one can obtain relevant information on the boundary of the spectral domain also in the presence of nontrivial gaugefields 
by monitoring the behavior of polynomials in the Wilson operator applied on random vectors.}

\FullConference{The XXV International Symposium on Lattice Field Theory\\
                 July 30 - August 4 2007\\
                 Regensburg, Germany}

\begin{document}

\section{Introduction}
Inverting the Wilson-Dirac operator is one of the dominant costs in simulating lattice QCD with dynamical fermions.  A possibility to avoid the explicit
inversion is to replace it by a polynomial approximation of the inverse Wilson-Dirac operator.\cite{dFT1996,FJ1997}  This allows moreover 
for a deviation from importance sampling with the Boltzmann factor to be compensated by reweighting, which can be useful. 
In addition simulations with odd numbers of flavors become possible.  Our focus here is the approximate inversion of the non-hermitian Wilson-Dirac operator. 
According to reference \cite{dFG1996} the approximation 
of the non-hermitian operator is supposed to be superior to the hermitian version.  Each approximation depends on the typical spectrum of the 
operator to be approximated.  Therefore, we here report about some spectral properties of the Wilson-Dirac operator with respect to Schr{\"o}dinger 
Functional(SF)\cite{LNWW1992} boundary conditions (BC) and investigate how they affect the polynomial approximation.

\section{Computing the spectrum semianalytically}

In the hopping parameter representation Wilson's fermion action \cite{W1974} reads
\begin{align}
S_f =\sum_{xy} \bar \psi(x) \left[\delta_{xy}-\kappa H_{xy} \right]\psi(y),
\end{align}
and the Wilson-Dirac operator is defined by $M_{xy}=\delta_{xy}-\kappa H_{xy}$, where the nearest neighbor interaction is carried by the hopping operator
\begin{align}
H_{xy}  &= \sum_{\mu=0}^3 \Big(  U_\mu(x)(1-\gamma_\mu)\delta_{x+\hat\mu,y}  + 
                                 U^\dagger_\mu(y)(1+\gamma_\mu)\delta_{x-\hat\mu,y}   \Big). \label{Hop}
\end{align}
Important spectral features of $H$ depend on the boundary conditions. 
In general $H$ is not normal i.e.~$[H,H^\dagger]\ne 0$.

In order to get a first idea of the spectrum and to develop methods for the general case we start by computing the eigenvalues of $H$ in the free case
i.e.~$U_\mu\equiv 1$ in the SF with vanishing background field.
Since translation invariance in the spatial directions still holds
we set 
\begin{align}
\psi(x) = \psi(x_0)\cdot \e^{i\vec p \vec x}.
\end{align}
using plain waves for the space dependence. This ansatz leads to a reduced 1-dimensional operator 
of the form
\begin{align}
E = \frac{1-\gamma_0}{2}\, \h + \frac{1+\gamma_0}{2}\, \hd +  i \gamma_1 \alpha,
\quad \alpha^2 = \sum_{k=1}^3 \sin^2(p_k) \label{RedOp}
\end{align}
acting on $\psi_0$.  In (\ref{RedOp}) $\h$ is the 1-dimensional hopping operator given 
for the SF by the nilpotent matrix
\begin{align}
 \h = 
\left( \begin{smallmatrix}
0&1&0&\cdots&0 \\[-2mm]
0&0&\ddots&&\vdots \\[-2mm] 
\vdots&&\ddots& 1&0 \\
0&&\cdots&0&1 \\
0&&\cdots&0&0 
\end{smallmatrix} \right)
.\end{align}
By itself it has only one $(T-1)$ fold degenerate zero eigenvalue. The matrix $E$ is related to $H$ by
\begin{align}
\txt{spec}(H)  = 2 \,\txt{spec}(E) + 2 \sum_{k=1}^3 \cos(p_k).
\end{align}
Finding the eigenvalues of $E$ is equivalent to locating the zeros of a smaller determinant
\begin{align}
0 = \det\left[\lambda^2_0 + \lambda^{\phantom{2}}_0 (\h + \hd) + \hd \h + \alpha^2\right].
\end{align}
We could not obtain them in closed form, but approximations are possible.
Here we simply compute the
eigenvalues of $E$ numerically for some range of $\alpha$.

\begin{figure}[ht]
\begin{minipage}{0.46\textwidth}
\hspace*{-0.3cm}
\includegraphics[scale=0.5]{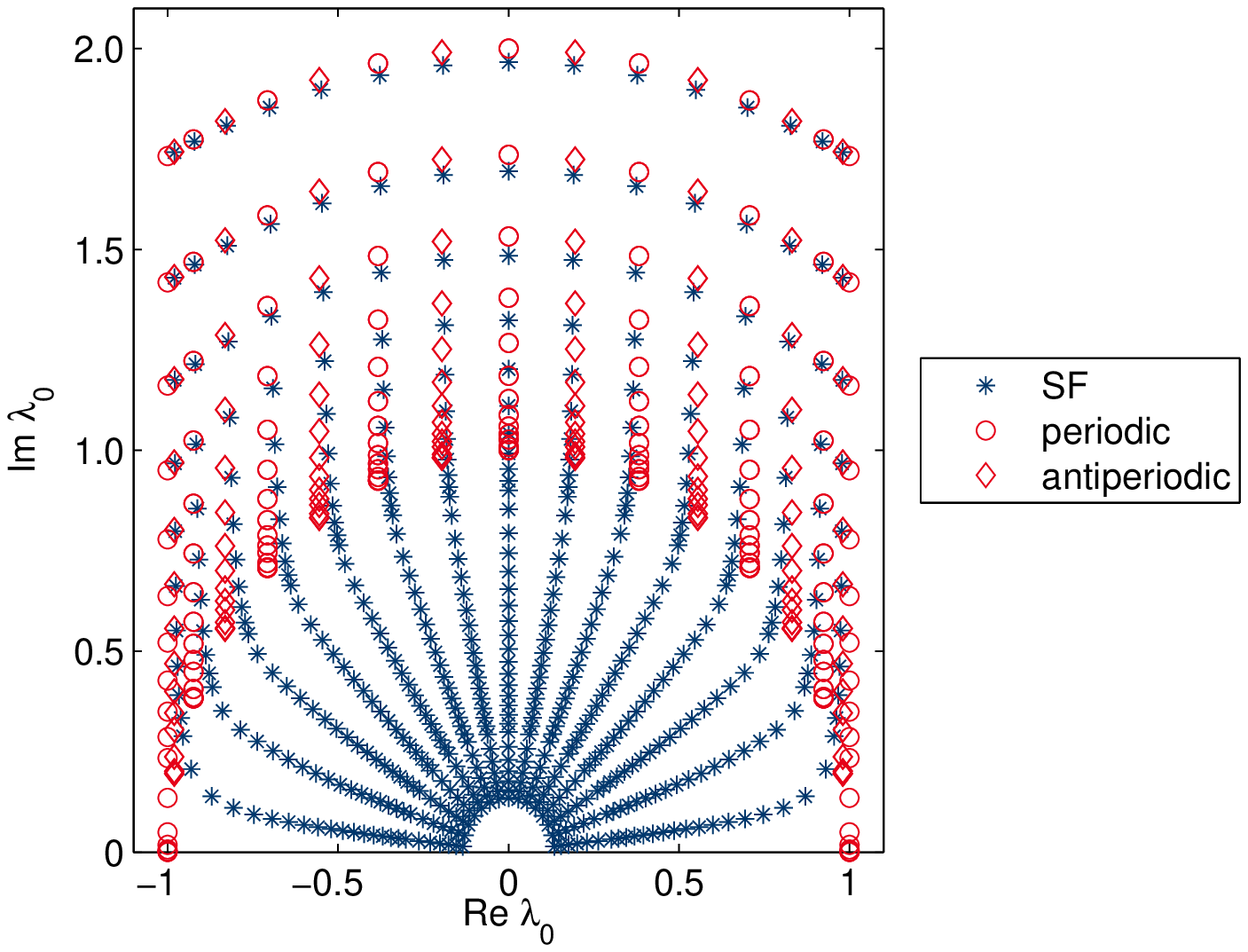}
\caption{Numerical spectrum of $E$ for $T=16$ and $\alpha^2 = \txt{tiny} \dots 3$.}
\label{SpectrumHeffSF}
\end{minipage}
\hfill
\begin{minipage}{0.51\textwidth}
\hspace*{-0.4cm}
\includegraphics[scale=0.51]{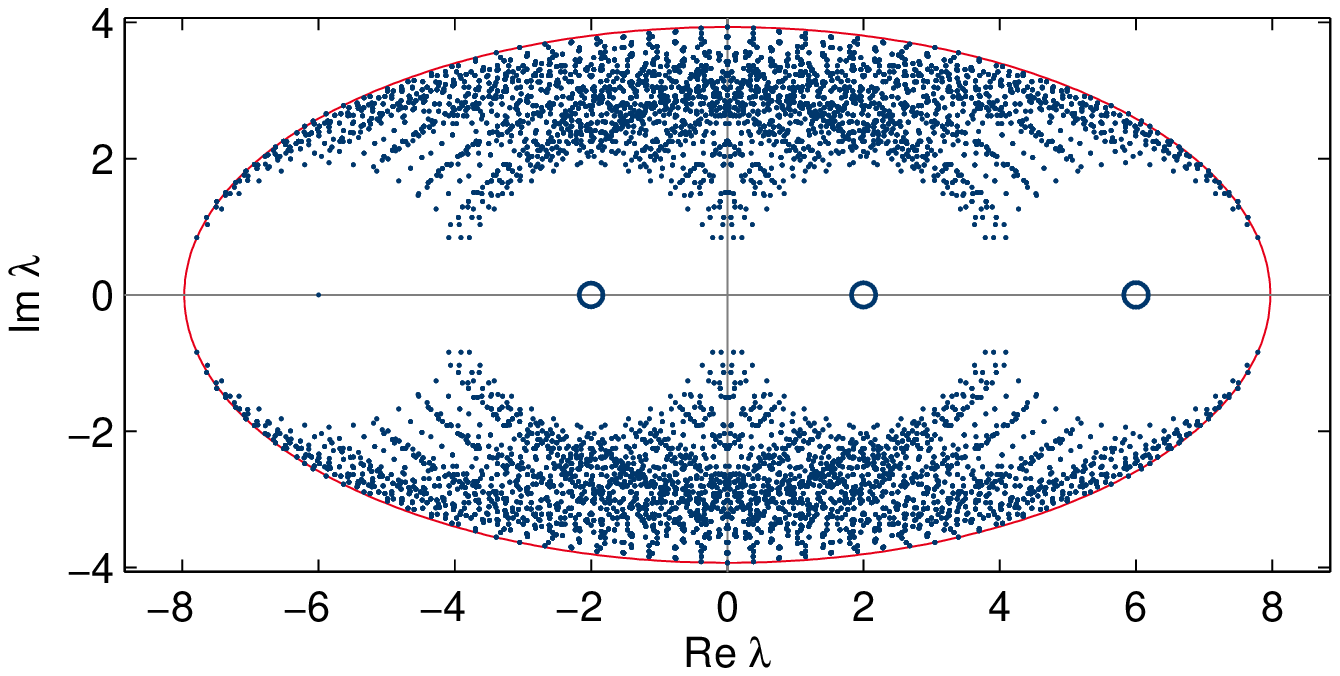}
\caption{Spectrum of the hopping operator $H$ on a $16^4$ lattice with $U_\mu \equiv \one$ and  SF boundary conditions.}
\label{SpectrumSF}
\end{minipage}
\end{figure}

These eigenvalues are shown in Fig.~\ref{SpectrumHeffSF} together with the corresponding ones for
(anti)periodic boundary conditions that follow
trivially from Fourier expansion. For small $\alpha$ the latter approach the value $1$
leading to zeromodes in $1-\kappa_c H$ (up to lattice artefacts for antiperiodic BC), where $\kappa_c$ equals $1/8$.
In the SF we see that the eigenvalues are `deflected' away from unity. For very small $\alpha$ one can show the behavior
\begin{align}
\abs{\lambda_0} \propto \alpha^{1/T}\quad \txt{as}\quad \alpha \to 0.
\label{asalpha}
\end{align}
This is how a gap of
order $1/T$ is maintained in the SF (even in the continuum limit).

We conclude this subsection by showing in Fig.~\ref{SpectrumSF} a complete spectrum of $H$ computed numerically.
An ellipse with major half axis $a=7.971$, minor half axis $b=3.932$ and eccentricity $e=\sqrt{a^2-b^2} = 6.933$ is drawn that obviously describes the spectral boundary
very well. In the centers of the void areas there are (degenerate) eigenvalues for which $\alpha$ vanishes due to the zero mode of $h_0$. Roundoff errors
in the eigenvalue routine in combination with the behavior (\ref{asalpha}) `inflate' three of these dots to small circles. 

\newpage
\section{Approximation of the inverse Wilson-Dirac operator}

To start simple we first approximate the inverse Wilson-Dirac operator by a geometric series
\begin{align}
M^{-1} \approx P^G_n(M) = \sum_{j=0}^n (\kappa H)^j = \frac{1-(\kappa H)^{n+1}}{1-(\kappa H)}, \quad \text{for}\;\; \norm{\kappa H} < 1,
\end{align}
and  define the remainder 
\begin{align}
R_{n+1} =  \one - M P^G_n(M) = (\kappa H)^{n+1}.
\end{align}
By construction $R_{n+1}$ is a small quantity and vanishes in the limit of $n\to\infty$.  Moreover, it allows for a recursive implementation and its convergence can be easily monitored by computing $\norm{R_{n+1} \eta}$, where $\eta$ is a Gaussian random vector normalized to $1$.  Approximating $M^{-1}$ by a geometric series requires a circular bound on the spectrum of radius $r = \kappa\abs{\lambda_\txt{max}(H)}<1$.\footnote{\label{f1}For non-diagonalizable matrices the same behavior is true asymptotically. This can be shown with the help of the Schur-decomposition.} From the spectral radius $r$ follows the rate of convergence
\begin{align}
\mu^G(\kappa) = - \ln (\kappa \vert \lambda_\txt{max}(H) \vert). \label{MuGeo}
\end{align}

As we have seen in the previous section the shape of the spectrum is elliptical.  This fact can be exploited to improve our approximation.  Expressing the remainder $R_{n+1}$ in terms of scaled and translated Chebyshev polynomials $T_n$\cite{M1977} we can derive an improved, recursive description, where only the eccentricity $e$ as elliptical parameter enters
\begin{align}
R_{n+1}(M) &= \frac{T_{n+1}((\kappa H)/e)}{T_{n+1}(1/e)}
= a_n \kappa H R_n(M) + (1-a_n)R_{n-1} (M) \label{RCheby}
\end{align}
with $R_1(M) = \kappa H$; $R_0(M) = 1$; $a_n=\left(1- a_{n-1}\,e^2/4 \right)^{-1}$ and $a_1 = \left(1- e^2/2\right)^{-1}$. The second equality in (\ref{RCheby}) follows from the recurrence relation of the Chebyshev Polynomials, $T_{n+1}(z)=2zT_{n}(z)-T_{n-1}(z)$. By virtue of the defining relation for the remainder we obtain also a recursive expression for the Chebyshev approximation of $M^{-1}$ 
\begin{align}
P_n^C(M) &= a_n(1+ \kappa H P_{n-1}(M)) + (1-a_n) P_{n-2}(M) 
\end{align}
with $P_1(M) = a_1(1+\kappa H)$ and $P_0(M) = 1$.  The rate of convergence in the limit of $n \to \infty$ follows using the identity $T_n(z) = \cosh(n\arcosh(z))$ and by replacing $\kappa H$ by its eigenvalues (cf.~footnote \ref{f1}) \cite{B1997}. The rate $\mu^C$ depends on the elliptical parameters $a$ and $e$ which themselves are proportional to $\kappa$
\begin{align}
\mu^C(a,e) = \ln\left(\frac{1+\sqrt{1-e^2}}{a+\sqrt{a^2 - e^2}}\right).
\label{Mu_Cheby}
\end{align}
For periodic BC the extent of the ellipse is known in the free case ($a=8\kappa$, $e=\sqrt{48}\kappa$) and thus $\mu^C$ becomes a function of $\kappa$ only,
$\mu^C(\kappa) = \ln\left( {(1+\sqrt{1-48\kappa^2})}/{(12 \kappa)}\right)$, and vanishes like (\ref{MuGeo}) for $\kappa \to \kappa_c$.

\section{Numerical results}
\begin{figure}[ht]
\begin{minipage}{0.51\textwidth}
\includegraphics[scale=0.5]{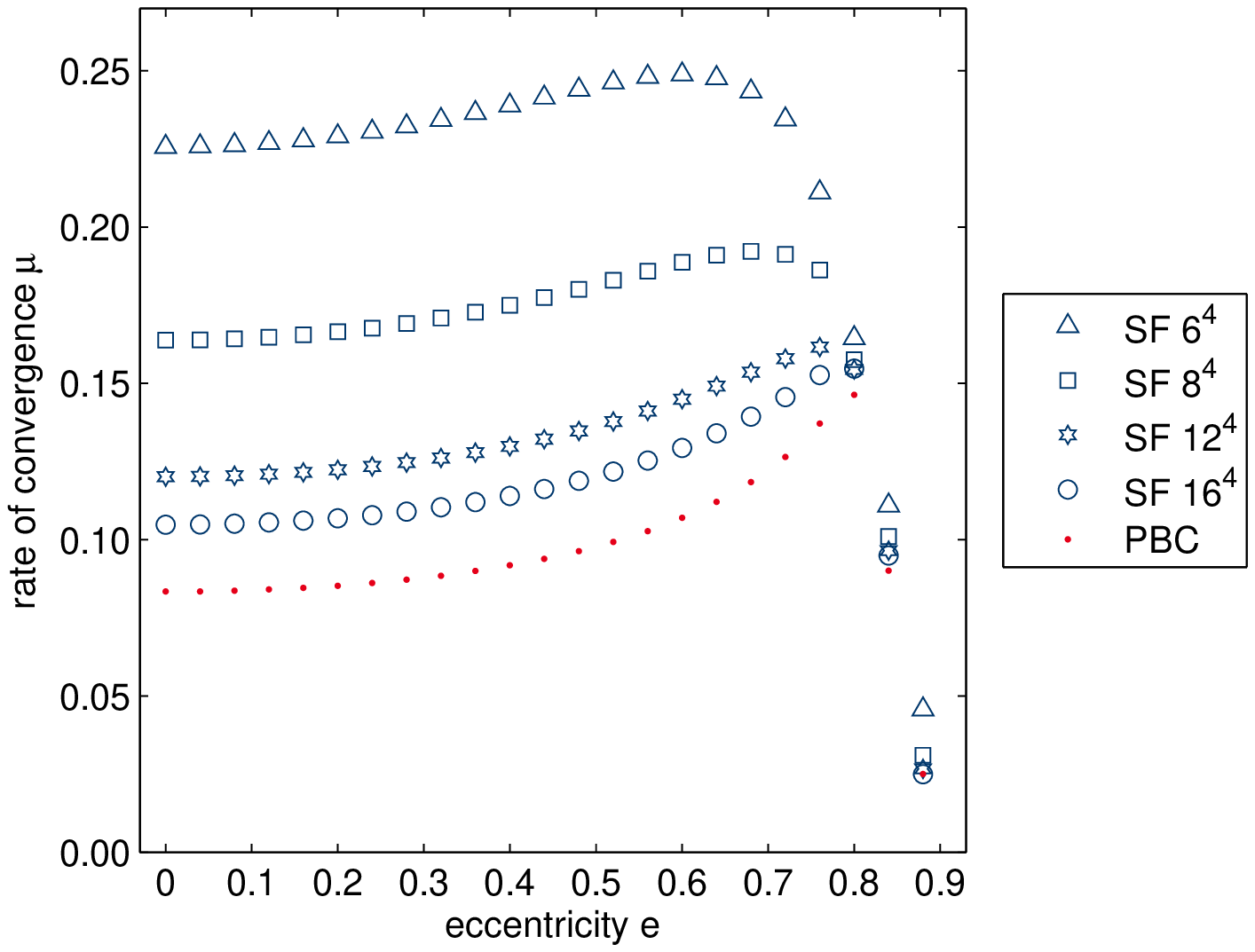}
\caption{Rate of convergence $\mu$ as function of the eccentricity $e$. $U_\mu \equiv \one$, $\kappa=0.115$ and each point is determined after $n=400$ iterations.}
\label{Mu_Ecc}
\end{minipage}
\hfill
\begin{minipage}{0.46\textwidth}
\hspace*{-0.4cm}
\includegraphics[scale=0.5]{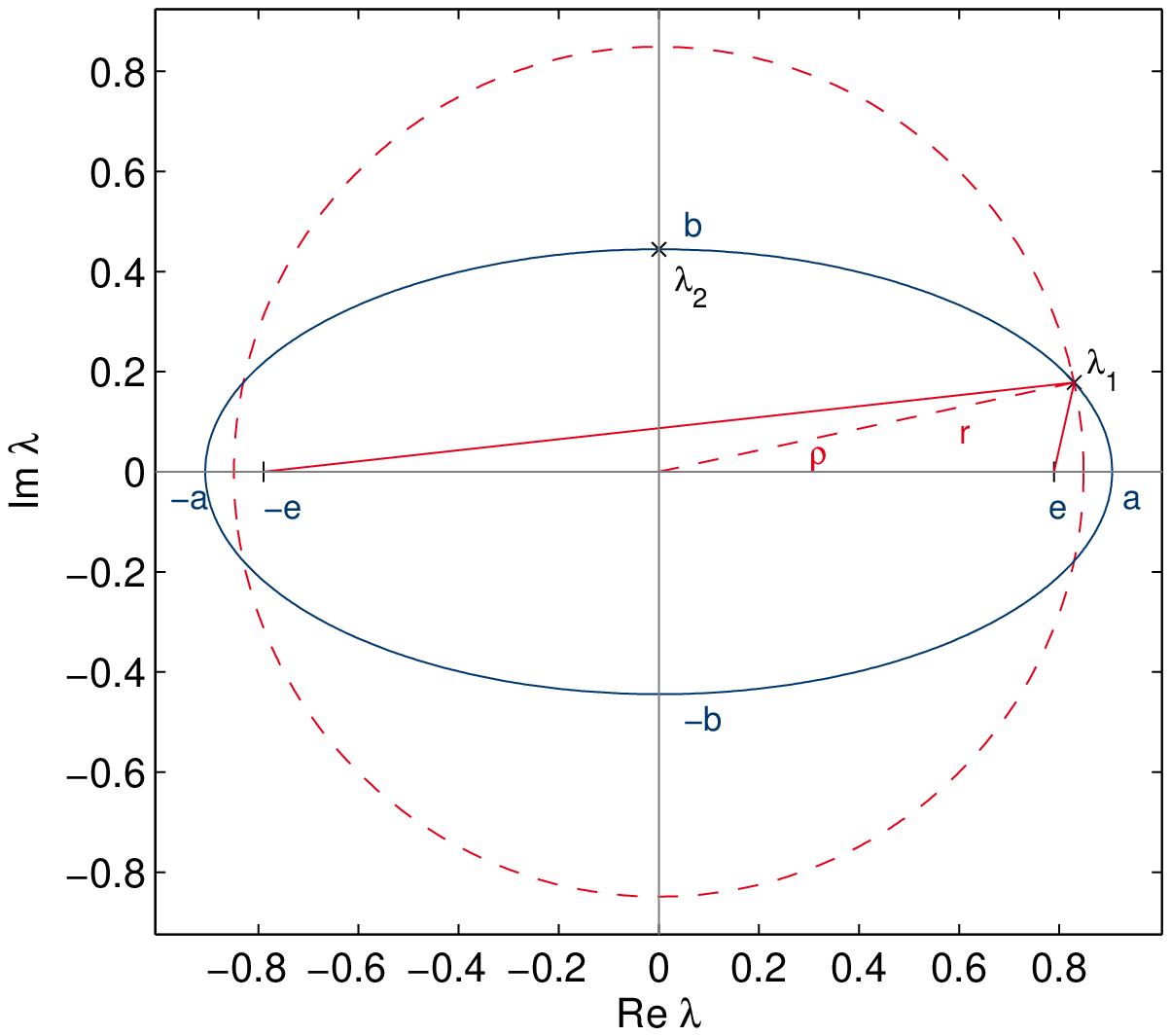}
\caption{Sketch how to get a good guess on the eccentricity starting from the largest eigenvalues.\newline}
\label{Ellipse}
\end{minipage}
\end{figure}

To test our approximations numerically we monitor the norm of the remainder as a function of $n$ and determine the convergence rate from the exponential decay.  We perform this test choosing $n=400$ and varying $e$ to obtain a scan over the eccentricity.  In Fig.~\ref{Mu_Ecc} the results are presented for various lattice sizes. 

Obviously, there is a dependence on the lattice size in the case of the Schr{\"o}dinger functional and for larger lattices the rate of convergence approaches the value corresponding to periodic BC which is independent of $L$ (red dots in Fig.~\ref{Mu_Ecc}).  Since the rate of convergence increases roughly by a factor $2$ from a circular bound ($e=0$, geometric series) to the optimal eccentricity, it is important to find this optimal value.  Therefore we look again at the spectrum of the Wilson-Dirac operator and focus our attention especially on the eigenvalue $\lambda_1$ with largest real part and $\lambda_2$ with largest imaginary part of $H$ as indicated in Fig.~\ref{Ellipse} (multiplied by $\kappa=0.115$). 

One way to obtain a guess on the eccentricity $e$ is to use the norm of $\lambda_2$ as value for the minor half axis $b$.  We are then seeking the ellipse which also passes through $\lambda_1$.  By the parameter form of an ellipse,
$
x= a\, \cos \rho; \; y=b\,\sin \rho
$,
and using $x + iy = \lambda_1 $ we find the major half axis
\begin{align}
\quad a = \txt{Re}\{\lambda_1\}/\cos(\rho) \quad \txt{with}\quad \rho = \txt{arcsin}(\txt{Im}\{\lambda_1\}/b).
\end{align}
Thus we can determine $e = \sqrt{a^2 - b^2}$.  Beside yielding a guess on $e$ we can moreover obtain an estimate on $\mu^C$ by eq.~(\ref{Mu_Cheby}) in this way.

There exist different methods to determine $e$.  In practice we are seeking a good guess on $e$  such that the convergence rate is high and its determination is easy.  These properties hopefully carry over when including a non-trivial gauge field. There we hope to find an eccentricity that changes only weakly between different gauge fields at fixed $\beta$ and $\kappa$. \\

For a first experiment with a non-trivial gauge field we start by generating 50 pure-gauge configurations on an $8^4$ lattice at $\beta = 6.0$ employing a Cabbibo-Marinari update \cite{CM1982}. Reading these configuration with \texttt{MATLAB} (version 7.3) and using its implementation of the Arnoldi algorithm we try to compute $\lambda_1$ and $\lambda_2$ on each configuration. Unfortunately, the algorithm converged only on a subset of the configurations.  Hence the mean values presented in Tab.~\ref{Exp_Mu} are just a rough estimation and within the quoted errors no dependence on the configuration is seen. 
\begin{table}[ht]
\centering
\begin{minipage}{14.6cm}
\renewcommand{\arraystretch}{1.3}
\begin{tabular}{||c|c|c||l|l||l|l|l|l||} \hline \hline
&$U_\mu$ &$\kappa$ &~~~~~$r $&~~~~~$\mu^G$&~~~~~$a$ &~~~~~$b$ &~~~~~$e$&~~~~~$\mu^C$ \\ \hline \hline
SF&$\one$& 0.115 &0.8489 &0.1638 &0.9060 &0.4444 & 0.7895 & 0.1781 \\ \hline
SF& $\beta=6.0$&0.135 &0.838(4)&0.1770(7)&0.843(5)& 0.47(1)&0.701(7)&0.268(3)\\ \hline\hline
P& $\one$ &0.115& 0.9200 &0.0834 & 0.9200 & 0.4600 & 0.7967 & 0.1506 \\ \hline
P& $\beta=6.0$ &0.135&0.865(5)&0.1456(8)&0.869(6)& 0.48(1)& 0.725(2)& 0.226(9)\\ \hline\hline
\end{tabular}
\caption{Expected values for $\mu$ and $e$ derived from measured maximal eigenvalues. SF Schr{\"o}dinger functional, P periodic boundary conditions.}
\label{Exp_Mu}
\end{minipage}
\end{table}
\renewcommand{\arraystretch}{1.0}

The results indicate that the spectrum of $H$ for non-trivial gauge fields is expected to be ``rounder'' and enclosed in an elliptical disc of smaller area than the one of the trivial gauge field.  We check our expectation by computing the polynomial remainder using the above determined $e$ in case of the Chebyshev approximation.  While for the trivial gauge field we find perfect agreement of both methods, the differences in the rate of convergence are larger for non-trivial gauge fields when using Chebyshev polynomials
\begin{table}[h]
\centering
\begin{minipage}{7.6cm}
\centering
\renewcommand{\arraystretch}{1.3}
\begin{tabular}{||c||l||l|l||}\hline \hline
  &~$\mu^G$ &$e$&~$\mu^C$  \\ \hline\hline
SF&~0.1776(3)~&0.701 &0.2127(2)\\\hline
P&~0.1482(4)~ &0.725 &0.1914(2)\\ \hline \hline
\end{tabular}
\caption{Computing the convergence from the remainder test as a measure on the ``integrated spectrum''.}
\label{Num_Mu}
\renewcommand{\arraystretch}{1.0}
\end{minipage}
\end{table}

\begin{figure}[ht]
\centering
\begin{minipage}{9.6cm}
\hspace*{-0.5cm}
\includegraphics[scale=0.67]{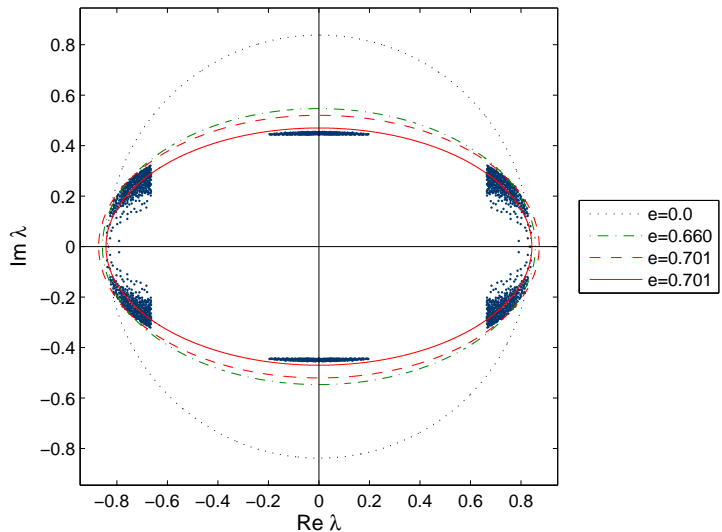}
\caption{Computing 2400 eigenvalues of $\kappa H$ on one gauge configuration at $\beta =6.0$ and $\kappa =0.135$ with SF boundary conditions.}
\label{Spec2400}
\end{minipage}
\end{figure}

To get a better understanding of this situation we computed for one SF configuration 800 eigenvalues with largest/smallest real part and 400 eigenvalues with largest/smallest imaginary part again using \texttt{MATLAB}.  Figure \ref{Spec2400} shows these data points in blue and the solid red line is the ellipse ($e = 0.701$, $a=0.843$) derived from the eigenvalue computation.  The predicted value for $\mu_C$ disagrees because the shape of the spectral boundary is not elliptical.  The circular bound is still estimated correctly as can be seen by the dotted black circle with radius $r=0.838$. Hence $\mu^G$ is in agreement. To illustrate our approximation using Chebyshev polynomials we note that by specifying $e$ a family of confocal ellipses is determined. From this family the ellipse of smallest extent which encloses \emph{all} eigenvalues specifies $a$ and $b$, which enter into (\ref{Mu_Cheby}). Hence the dashed red ellipse in Fig.~\ref{Spec2400} represents better the one corresponding to the Chebyshev approximation. Here $a=0.872$ and we compute $\mu^C=0.208$. 

 Increasing $a$ to $0.857$ and $b$ to $0.547$ we yield a different ellipse ($e=0.660$) shown with dash-dotted green line.  This one encloses the computed spectrum even better and leads to the prediction $\mu^C(e=0.660) \approx 0.221$.  Taking this smaller value of $e$ as input for the Chebyshev approximation we find for the rate of convergence $\mu^C= 0.2207(2)$ confirming the predicted value.

\section{Conclusion and outlook}
These preliminary studies show that a good understanding of the structure of the spectrum seems to be important to implement an algorithm approximating the inverse Wilson-Dirac operator with good performance.  Moreover, useful information on how to tune such an algorithm is obtained.  

Probably, the deviation of the spectrum in the SF from an elliptic disk is an artefact of small lattice sizes. A check with e.g.~a $12^4$ lattice would be desirable but seems to be numerically challenging. Moreover we like to study the effect of $O(a)$ improvement (Sheikholeslami-Wohlert term) and the effects of preconditioning.

\paragraph{Acknowledgement\\}
This work is supported by the DFG within the SFB/TR 9 ``Computational Particle Physics''.

\end{document}